# Precise Measurements of the Decay of Free Neutrons

Dirk Dubbers[1] and Bastian Märkisch[2]

[1]Physikalisches Institut, Universität Heidelberg, Im Neuenheimer Feld 226, 69120 Heidelberg, Germany

[2]Physik-Department, Technische Universität München, James-Franck-Straße 1, 85748 Garching, Germany

**Abstract:** The impact of new and highly precise neutron $\beta$ decay data is reviewed. We focus on recent results from neutron lifetime, $\beta$ asymmetry, and electron-neutrino correlation experiments. From these results, weak interaction parameters are extracted with unprecedented precision, possible also due to progress in effective field theory and lattice QCD. Limits on New Physics beyond the Standard Model derived from neutron decay data are sharper than those derived from high-energy experiments, except for processes involving right-handed neutrinos.

# Contents

# 1. Introduction

The $\beta$ decay of the neutron into proton, electron and electron-antineutrino $n \to p^+ e^- \bar{\nu}_e$ is a fascinating field of study. It addresses many basic issues of contemporary physics, relevant at the smallest scales of elementary particle physics up to the largest scales of space-time in cosmology, astrophysics, and solar physics. Most experiments in nuclear and particle physics use accelerators with energies from MeV up to the TeV scale. In contrast, slow, i.e., thermal, cold, or ultracold neutrons (UCN), see left side of Table 1 below, have energies from meV down to peV, 24 orders of magnitude below the TeV scale, which is a world apart, unfamiliar to the majority of nuclear or particle physicists.

Due to their neutrality, neutrons cannot easily be accelerated, deflected, or focused. This is the burden of experimental neutron physics, but at the same time, it is its main asset: Neutrons react to all known forces of nature but not to the usually overwhelming electrostatic force. This makes neutrons highly sensitive to the subtlest effects.

In Section 2 we first want to convey the special flavor of experimental work with slow neutrons. The theory of neutron decay, including effective field theory (EFT) and lattice gauge theory, is treated in Section 3. Recent neutron decay experiments are reviewed in Sections 4. Section 5 discusses applications of neutron decay data. EFT and lattice gauge theory make it possible to compare limits on New Physics obtained from low and high-energy experiments, as shown in Section 6.

Previous reviews on neutron decay include Refs. (1-5), a series of articles introduced by Ref. (6), plus a recent neutron conference (7). For reviews on beta decay in general, see (8-10), and references therein. In view of these previous reports, the present review will focus on the exciting developments of the past few years.

Neutron $\beta$ decay is a subfield of neutron particle physics, which covers in addition searches for a neutron electric dipole moment (EDM) (11), of neutron-antineutron (12) and neutron-mirror neutron oscillations (13), all sensitive to neutron energy shifts on the $10^{-24}$ eV scale. Furthermore, there exist studies on hadronic parity violation (14), on gravitational quantum states and limits on dark energy (15), and on the foundations of quantum mechanics via neutron interferometry (16), together with firsthand observations of geometric phases (17) and dressed particle effects (18).

# 2. Working with Slow Neutrons

The neutron sources used for $\beta$ decay experiments are of two types: Reactor sources like FRM II, ILL, NIST, PNPI, TRIGA-Mainz, and spallation sources like ESS, J-PARC, LANL, PSI, SNS. (For acronyms, see Terms and Definitions at the end of this text.) We first list some of the peculiarities of cold and ultracold neutrons.

## 2.1. Generation, Transport, and Storage of Neutrons

**2.1.1. Generation of cold and ultracold neutrons.** Reactor sources are conventional nuclear reactors running at low power between 20 MW and 100 MW, optimized for a continuous high neutron flux. Spallation sources use accelerators to bombard a target of heavy nuclei with protons of typically GeV energy, releasing ~30 neutrons per collision. In both cases, the neutrons, which are generated with energies of up to several hundreds of MeV, must first be moderated to room temperature, usually in a bath of heavy water several meters in diameter. Neutrons can be cooled further down to ~30 K in a secondary moderator containing several liters of liquid deuterium, or another material of low capture cross section. Ultracold neutrons, first obtained in in the late 1950s (19, 20), are nowadays produced by reflecting neutrons from the receding blades of a turbine (21), or by having neutrons lose their energy to local excitations in superfluid helium or in solid deuterium (22). Comparative measurements on different UCN sources are reported in (23).

**Table 1. Mean kinetic variables (left) and potential energies (right) of slow neutrons***

| Neutron | $E_0$ | $v_0$ | $T$ |
|---|---|---|---|
| Thermal | 25 meV | 2200 m/s | 273 K |
| Cold | 2.5 meV | 700 m/s | 24 K |
| Ultracold | < 300 neV | < 8 m/s | $T_{\text{eff}} \approx 3$ mK |

| Interaction | Potential |
|---|---|
| Magnetic | 60 neV/T |
| Gravity | 102 neV/m |
| Strong | <300 neV |

* $E_0$, kinetic energy at most probable velocity $v_0$ at temperature $T$.

**2.1.2. Transport of neutrons.** Slow neutrons diffusing out of the moderator into beam tubes behave more like a gas than a beam of particles. They can only be handled via their rather feeble magnetic, gravitational, and strong interactions, see the right side of **Table 1**. In condensed matter, the de Broglie wavelength of slow neutrons are comparable to the interatomic distances (~Å), while the range of the strong force is limited to the size of the nuclei (~fm). Therefore, in ordinary matter the neutron's strong interaction (~$10^8$ eV) is suppressed by a factor of

~(1 fm/1 Å)$^3$ = 10$^{-15}$ to at most 300 neV. This small interaction, which is repulsive for most materials, is at the basis of neutron guides, which are glass channels of up to 20 cm height by 6 cm width and 160 m length, whose inner surfaces are coated with a totally reflecting layer of metal (24). For cold neutrons, the critical angle of total reflection $\theta_c \approx (E_{\text{pot}}/E_{\text{kin}})^{1/2}$ can reach ~10 mm/m. The first "ballistic" neutron guide (25, 26) – of low loss due to a guide's cross section that varies along the beam axis – was coated with a multilayer "supermirror" (27) and was dedicated to neutron-particle physics. Monochromatic neutrons are obtained with fast-turning mechanical velocity selectors, or by time-of-flight if the beam is pulsed.

**2.1.3. Storage of ultracold neutrons.** In a guide for UCN, the repulsive potential exceeds the kinetic energy of the neutrons, hence, they are totally reflected for any angle of incidence. If the UCN fill a totally reflecting reservoir at the end of a guide, of volume up to 1 m$^3$, whose entrance door is closed after filling, then this reservoir acts as a neutron bottle where neutrons can be stored for periods much longer than their lifetime of ~15 minutes. In the bottle the UCN follow flight parabolas, interrupted only by elastic reflections from the walls. The relatively high temperature of the material walls does not heat the UCN because phonon densities are low in condensed matter, for tiny residual interactions, see (28). In the vertical direction, UCN can be confined gravitationally if the height of the bottle is of the order of meters. Polarized UCN can also be stored magnetically in a repulsive magnetic potential when field maxima are of order Tesla. For details of UCN storage, see (29-31).

## 2.2. Polarization and Detection of Neutrons

**2.2.1. Spin-polarization of neutrons.** Thermal or cold neutrons can be polarized via the spin-dependence of their magnetic or strong interaction. This is achieved by total reflection from magnetized supermirrors (32), or by transmission through nuclear spin-polarized $^3$He gas, achieving >99% polarization with 10$^{-4}$ precision for beams of up to decimeter diameter, see (33, 34). Neutron spins are inverted in-flight with near 100% efficiency by the adiabatic fast-passage method. To polarize a beam of UCN, one simply blocks one spin component by a magnetic barrier.

**2.2.2. Detection of neutrons.** Slow neutrons can be only detected destructively, via various neutron capture reactions. For a long time, $^3$He filled Geiger-Müller tubes were standard neutron detectors. Due to the $^3$He-shortage of recent years, several new neutron detection systems were developed, see (35, 36). To detect UCN, one must let them fall down by ~1 m to gain sufficient energy to overcome the reflective potential of the detector surface.

# 3. Theory of Neutron Decay

## 3.1. V – A Theory

**3.1.1. The electroweak interaction.** In the standard model (SM), the electromagnetic and the weak interactions are unified at the electroweak scale $v = 2M_W c^2/g = 246$ GeV, $\sqrt{2}$ times the expectation value of the Higgs field. $M_W = 80.4$ GeV$/c^2$ is the mass of the $W^\pm$ boson, the mediator of weak transitions. The dimensionless weak coupling constant $g$ is related to the Fermi coupling constant by $G_F/(\hbar c)^3 = (\sqrt{2}/8)g^2/(M_W c^2)^2 = 1.166 \times 10^{-5}$ GeV$^{-2}$, known with great precision from muon decay (37).

Symmetry requirements tell us much about the interaction. The Lagrangian must be a scalar under Lorentz transformations. For a four-fermion "contact" interaction at a single point of space-time, such as $\beta$ decay with energy $E \ll m_W c^2$, the operators in the Lagrangian are the scalar products of five covariant bilinear objects, listed in **Table 2**. These objects involve the 4×4 $\gamma$ matrices, the building blocks of relativistic Dirac theory, and the 4-component fermionic Dirac spinors $\psi$.

**Table 2. The covariant bilinears and their properties under Lorentz transformation***

| Covariant bilinear | $\bar{\psi}\psi$ | $\bar{\psi}\gamma_\mu\psi$ | $\bar{\psi}\gamma_\mu\gamma_5\psi$ | $\bar{\psi}\sigma_{\mu\nu}\psi$ | $\bar{\psi}\gamma_5\psi$ |
|---|---|---|---|---|---|
| Transforms as | Scalar S | Vector V | Axial vector A | Tensor T | Pseudoscalar P |
| No. of components | 1 | 4 | 4 | 6 | 1 |

*$\sigma_{\mu\nu} = \frac{1}{2}i(\gamma_\mu\gamma_\nu - \gamma_\nu\gamma_\mu)$, $\gamma_5 = i\gamma_0\gamma_1\gamma_2\gamma_3$

For unknown reasons, only the V – A linear combination $\bar{\psi}\gamma_\mu\psi - \bar{\psi}\gamma_\mu\gamma_5\psi = \bar{\psi}\gamma_\mu(1-\gamma_5)\psi$ seems to be realized in nature. A parity transformation $(t, \mathbf{x}) \to (t, -\mathbf{x})$ changes the sign of $\bar{\psi}\gamma_\mu\psi$ but not that of $\bar{\psi}\gamma_\mu\gamma_5\psi$, while both operators occur with equal magnitude; hence the V – A coupling is maximally parity violating. The operator $1 - \gamma_5$ projects out left-handed neutrinos $\nu_L$ and right-handed anti-neutrinos $\bar{\nu}_R$.

**3.1.2. Neutron decay at the quark level.** On the quark level of neutron decay, $d \to ue\bar{\nu}_e$, with up and down quarks $u$ and $d$, the V – A Lagrangian density (shortly called Lagrangian hereafter) is

$$\mathcal{L}_q = -(G_F/\sqrt{2})\, V_{ud}[\bar{\psi}_e\gamma_\mu(1-\gamma_5)\psi_{\nu_e}][\bar{\psi}_u\gamma^\mu(1-\gamma_5)\psi_d)], \qquad 1.$$

with $G_F/\sqrt{2} = 0.634 \times 10^{-7}$ GeV fm$^3$, and summation over the repeated index $\mu$ = 0, 1, 2, 3. The components of the relativistic four-vectors $\psi$ are reshuffled by multiplication with the $\gamma$ matrices.

The unification of the weak and the electromagnetic interaction requires the weak vector current to be conserved in the same way as the electromagnetic vector current (CVC) is conserved. Weak CVC holds across the three particle families, but only if we take into account that, again for reasons unknown, the weak eigenstates of the quarks are slightly rotated away from their mass eigenstates, with angles given by the Cabibbo-Kobayashi-Maskawa (CKM) quark-mixing matrix (38, 39), whose first (up-down) matrix element $V_{ud}$ enters Equation 1. The CKM matrix must be unitary, which means $|V_{ud}|^2 + |V_{us}|^2 + |V_{ub}|^2 = 1$ for its first row, where $V_{us}$ describes the mixing of the $u$ quark with the strange $s$ quarks of the second particle family. The third term $|V_{ub}|^2$, which involves the bottom $b$ quarks of the third family, is negligible in this context.

**3.1.3. Neutron decay at the nucleon level.** Neutron decay is observed not on the quark level but on the nucleon level. The complicated internal structure of the nucleons must be accounted for by introducing form factors, with $\bar{\psi}_p\gamma_\mu[g_V(q^2) + g_A(q^2)\gamma_5]\psi_n$ replacing $\bar{\psi}_u\gamma_\mu(1-\gamma_5)\psi_d$. In neutron decay, the four-momentum transfer $q$ can be neglected so the form factors can be taken at $q^2 \to 0$, and are then called coupling constants. Analogous to the anomalous magnetic moment component of the electromagnetic current (with which it forms a weak isospin triplet), a term called weak magnetism (40) must be added, proportional to the difference of neutron and proton anomalous magnetic moments, $\kappa_p - \kappa_n = 3.7$ in nuclear magnetons, leading to a ~1% extra decay amplitude.

In contrast to the weak vector current, electroweak unification does not require the conservation of the weak axial-vector current, which has no analog in electromagnetism. In principle, the axial coupling $g_A$ could be very large in the strongly interacting environment of nucleons. Yet, all the complications due to the interior structure of the neutron can been reduced to a simple numerical factor $\lambda \equiv g_A/g_V \approx -1.275$, surprisingly close to $-1$. Although required by the partially conserved axial current hypothesis (PCAC) (41), an induced pseudoscalar term, though present, is still negligible in neutron decay, in spite of a large coupling $g_P \approx 230$ from Lattice QCD (42). With Hamiltonian density $\mathcal{H} = -\mathcal{L}$, the nucleon transition matrix element then is

$$M_{n\to p} = (G_F/\sqrt{2})\, V_{ud} \bar{\psi}_p[\gamma_\mu(1+\lambda\gamma_5) + \frac{\kappa_p-\kappa_n}{m_p+m_n}\sigma_{\mu\nu}q^\nu]\psi_n. \qquad 2.$$

In the decay probability, proportional to $|M_{n \to p}|^2 \rho$, the density of final states $\rho$ determines the spectra and angular distributions of the outgoing particles, for reviews see (43,44).

## 3.2. The observables

The neutron's $\beta$ spectrum, shown in **Figure 1a**, has an endpoint energy $E_0 = 0.782$ MeV, while the proton's endpoint energy is only 751 eV. Integration of the spectrum over electron energy $E$ gives the neutron decay rate

$$\tau_n^{-1} = \frac{G_F^2 (mc^2)^5}{2\pi^3 \hbar (\hbar c)^6} V_{ud}^2 (1 + 3\lambda^2) f (1 + \delta_R')(1 + \Delta_R), \qquad 3.$$

with a phase-space statistical factor $f$, and radiative corrections $\delta_R'$ and $\Delta_R$, to be explained in Section 5.2.1.

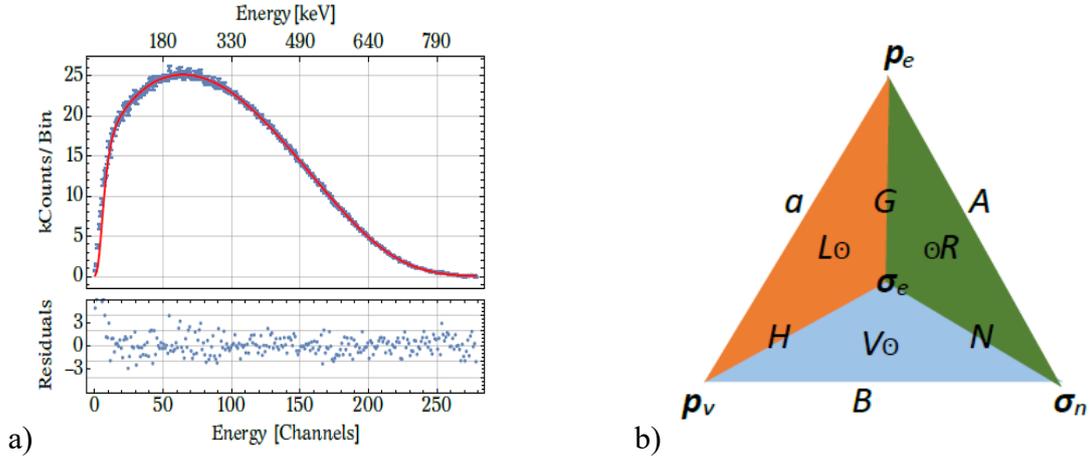

**Figure 1.** Observables in neutron decay. **a)** Measured and calculated $\beta$ energy spectrum, from Reference (50). **b)** Recipe for the construction of double and triple scalar products in neutron decay correlations. The experimentally accessible momentum and angular momentum vectors $\boldsymbol{p}_e$, $\boldsymbol{p}_{\bar{\nu}}$, $\boldsymbol{\sigma}_n$, $\boldsymbol{\sigma}_e$ are placed on the corners, their scalar products on the edges, and their time reversal $T$-violating triple products on the faces of a tetrahedron, shown as seen from above, see text.

Besides the neutron lifetime $\tau_n$, there are many other observables, which can be described by so-called correlation coefficients, see **Table 3**. The possible correlations are obtained by writing down all dimensionless scalars that can be constructed from the momenta $\boldsymbol{p}$ and angular momenta $\boldsymbol{\sigma}$ of the four particles involved in neutron decay. Slow neutrons are effectively at rest, $\boldsymbol{p}_n = 0$, and the polarization $\boldsymbol{\sigma}_p$ of the slow protons is practically inaccessible. The equally inaccessible neutrino momenta $\boldsymbol{p}_{\bar{\nu}}$ can be reconstructed from the measured electron and proton momenta. This leaves us with two vectors $\boldsymbol{p}_e, \boldsymbol{p}_{\bar{\nu}}$ and two axial vectors $\boldsymbol{\sigma}_n, \boldsymbol{\sigma}_e$. They all change sign under time reversal operation *T*, hence scalar products of an even number of these vectors are *T*-invariant, while products of an odd number of vectors signal *T*-violation.

**Table 3. Neutron *β* decay correlations measured up to now**

| Correlation | Scalar product | Symmetry violated* |
|---|---|---|
| Electron-antineutrino correlation *a* | $a\, \mathbf{p}_e \cdot \mathbf{p}_{\bar{\nu}}$ | – |
| Fierz term *b* | $b$ | SM: V–A |
| Beta asymmetry *A* | $A\, \mathbf{p}_e \cdot \boldsymbol{\sigma}_n$ | P |
| Neutrino asymmetry *B* | $B\, \mathbf{p}_{\bar{\nu}} \cdot \boldsymbol{\sigma}_n$ | P |
| Proton asymmetry *C* | $C\, \mathbf{p}_p \cdot \boldsymbol{\sigma}_n$ | P |
| Triple correlation *D* | $D\, \boldsymbol{\sigma}_n \cdot (\mathbf{p}_e \times \mathbf{p}_{\bar{\nu}})$ | T |
| Triple correlation *R* | $R\, \boldsymbol{\sigma}_e \cdot (\boldsymbol{\sigma}_n \times \mathbf{p}_e)$ | P, T |
| *n*-pol. – *e*-pol. correlation *N* | $N\, \boldsymbol{\sigma}_e \cdot \boldsymbol{\sigma}_n$ | – |

\* *P* = Parity, *T* = Time reversal

Six two-fold, four triple, four 4-fold, and one 5-fold scalar products can be formed from these vectors. Of the 16 possible correlation coefficients (including the Fierz term), Table 3 lists the 8 that have been studied so far, see (45), (46), (47) for their dependence on the weak couplings $g_i$, $i$ = V, A, S, T. In the SM, all correlation coefficients are functions solely of $\lambda = g_A/g_V$. Hence, in view of the many observables, the problem is overdetermined, and various tests beyond the SM are possible, as will be described in Sections 5 and 6. The Fierz interference term *b* (48,49) vanishes in the SM (more in Section 5.3.4).

**Figure 1b** is meant as a mnemonic for the six double correlations with coefficients *a*, *A*, *B*, *G*, *H*, *N* and the four *T* violating triple scalar correlations with coefficients *D*, *L*, *R*, *V* (*D* on the bottom face not shown). As an example, the $\mathbf{p}_e$-$\boldsymbol{\sigma}_n$ correlation gives the angular distribution of the *β* asymmetry, $1 + A\mathbf{p}_e \cdot \langle\boldsymbol{\sigma}_n\rangle c/W_e = \mathbf{1} - AP_n(v/c)\cos\theta$, with the angle $\theta$ between $\mathbf{p}_e$ and $\boldsymbol{\sigma}_n$, the electron helicity $cp_e/W_e = -v/c$ with total energy $W_e = E + mc^2$, and the neutron polarization $P_n = \langle\sigma_n\rangle$.

Neglecting small radiative and other corrections, the dependence of the *β*-*v* correlation *a*, the *β* asymmetry *A*, and the neutrino asymmetry *B*, to name just a few, on $\lambda = g_A/g_V$ is (45),

$$a = \frac{1-\lambda^2}{1+3\lambda^2}, \quad A = -\frac{2\lambda(\lambda+1)}{1+3\lambda^2}, \quad B = \frac{2\lambda(\lambda-1)}{1+3\lambda^2}. \qquad 4.$$

## 3.3. Effective Field Theory

The physics of elementary particles is based on quantum field theory (QFT). In QFT one first postulates a set of symmetries, which include the symmetries of the SM group SU(3) × SU(2) × U(1) and the Poincaré group of special relativity. Then one writes down the most general Lagrangian $\mathcal{L}_0$ that obeys these symmetries and that is renormalizable (has removable divergences). The elementary particles are defined and distinguished from each other solely by their quantum numbers under the postulated symmetries. For a given experimental energy scale $\Lambda_0$, the only fields which enter the Lagrangian $\mathcal{L}_0$ are those whose masses can be produced at this scale.

**3.3.1. Properties of effective field theories.** One often wants to know whether some virtual physical process, taking place at an inaccessible energy scale $\Lambda \gg \Lambda_0$, has a measurable effect in an experiment done at a low energy scale $\Lambda_0$. The aim of EFTs is to construct a low-energy effective Lagrangian $\mathcal{L}_{\text{eff}}$ that takes into account such virtual high-energy processes. To this end, one first writes down all possible operators $\mathcal{O}_i$ of higher dimensions that obey the same set of symmetries as the well-known Lagrangian $\mathcal{L}_0$ and are built from all known fields with masses up to the experimentally accessible scale $\Lambda_0$. Then one expands the Lagrangian in terms of these operators, usually restricting them to dimension 6,

$$\mathcal{L}_{\text{eff}} = \mathcal{L}_0 + \sum_i \frac{\varepsilon_i^{(6)}}{\Lambda^2} \mathcal{O}_i^{(6)}, \qquad 5.$$

with dimensionless so-called Wilson coefficients $\varepsilon_i$ (we drop the superscript), where $\Lambda$, which does not depend on $i$, sets the global energy scale. Operators of odd dimensionality violate the already strongly constrained baryon and/or lepton number conservation, while dimension 8 operators $\mathcal{O}_i^{(8)}$ are highly suppressed by their prefactor $1/\Lambda^4$. Results from EFTs are universal and model independent. With EFTs one can calculate the $\varepsilon_i$ once and for all for a given type of process. The EFT approach had to overcome many difficulties, as described in Reference (51). Several examples of successful EFTs are given in Reference (52).

How is the dimension of a field operator defined? With natural units $\hbar = c = 1$, the quantities $E$, $p$, and $m$ can be expressed in units of mass, with mass exponents +1, while phases like $Et$ and $px$ and are dimensionless, hence $t$ and $x$ have mass exponents –1. The mass exponent of an operator $\mathcal{O}_i$ is called its (mass-)dimension, which we denote as $\dim(\mathcal{O}_i)$. The action $S$ has the same dimension as Planck's constant $\hbar$, hence $\dim(S) = 0$ under $\hbar = 1$. With action $S = \int \mathcal{L}\, \mathrm{d}^4 x$ and $\dim(\mathrm{d}^4 x) = -4$, all Lagrangian densities have $\dim(\mathcal{L}) = 4$, for instance $\dim(\mathcal{L}_{\text{eff}}) = \dim(\mathcal{L}_0) = \dim\left(\mathcal{O}_i^{(6)}\right) + \dim(1/\Lambda^2) = 6 - 2 = 4$, see **Table 4**.

**Table 4. The dimensions of various field operators**

| Field | With $\dim(\mathcal{L}) = 4$ and: | follows for dim(Field) |
|---|---|---|
| Photon field tensor $F$ | $\mathcal{L} = -\tfrac{1}{4} F_{\mu\nu} F^{\mu\nu}$ | $\dim(F) = 2$ |
| with 4-vector potential $A$ | $F_{\mu\nu} = \partial_\mu A_\nu - \partial_\nu A_\mu$ | $\dim(A) = 1$ |
| Fermion field $\psi$ | $\mathcal{L} = -m\bar{\psi}\psi$ | $\dim(\psi) = 3/2$ |
| Gluon field $G$ | $\mathcal{L} = -\tfrac{1}{4} G_{\mu\nu} G^{\mu\nu}$ | $\dim(G) = 2$ |
| Higgs field $\varphi$ | $\mathcal{L} = -m^2 \varphi^\dagger \varphi$ | $\dim(\varphi) = 1$ |

**3.3.2. Effective field theory of neutron decay.** The conventional description of $\beta$ decay in Section 3.1 turns out to be a low-energy EFT-approximation to the SM. Indeed, the dimension of the operator $\mathcal{O}_i \propto [\bar{\psi}_e \gamma_\mu (1-\gamma_5)\psi_{\nu_e}][\bar{\psi}_u \gamma^\mu (1-\gamma_5)\psi_d]$ in Equation 1 is $4 \times \dim(\psi) = 6$ (from Table 3, or from $\psi \propto 1/\sqrt{V}$). The dimension of the prefactor $G_F$, in ordinary units GeV fm$^3$, is $\dim(G_F) = 1 - 3 = -2$. Therefore, $\mathcal{L}_q$ in Equation 1 is an effective Lagrangian of the same type as Equation 5, with prefactor $\varepsilon/\Lambda^2 = G_F/\sqrt{2}$ and a single dimensionless Wilson coefficient $\varepsilon$. Knowledge of $G_F$ gives no information on the weak interaction scale $\Lambda$.

To obtain the most general $\mathcal{L}_{\text{eff}}$ the V – A operators of Equation 1 must be complemented with all other possible four-fermion contact operators $\mathcal{O}_i$ of dimension 6. These are built from the complete set of $i$ = V, A, S, T, P-operators listed in Table 2, operating on the same fermion fields of $u, d, e, \bar{\nu}_e$ as $\mathcal{L}_q$. To give an example, the EFT tensor T operator is $\mathcal{O}_i \propto (\bar{\psi}_e \sigma_{\mu\nu} \psi_{\nu_e})(\bar{\psi}_u \sigma^{\mu\nu} \psi_d)$ with Wilson coefficients $\varepsilon_T$. These V, A, S, T, P-operators had already been included in Lee and Yang's seminal article on parity violation (53). A simple one-to-one correspondence exists between Lee and Yang's coefficients $C_i$, $C_i'$ and the Wilson coefficients $\varepsilon_i$. For our example of a tensor T interaction, this correspondence reads $C_T + C_T' = (8/\sqrt{2}) G_F V_{ud} g_T \varepsilon_T$. For a complete list, see (10).

One can also add to the left-handed (L = V–A) terms right-handed (R = V + A) terms, by replacing in the quark sector $1 - \gamma_5$ by $1 + \gamma_5$, and one preferably replaces the $i$ = V, A, S, T, P scheme by the $i$ = L, R, S, T, P scheme, which strongly reduces correlations between the various observables. One then speaks of left- and right-handed currents with Wilson coefficients $\varepsilon_L$ and $\varepsilon_R$. If, instead, $1 \mp \gamma_5$ appears in the leptonic sector, one refers to the left- and right-handed neutrinos, with Wilson coefficients $\tilde{\varepsilon}_i$ in the notation of Reference (10) and others.

In order to implement EFT at the nucleon level, one adds to Equation 2 all L, R, S, T, P nucleon matrix elements (54,55), like the above tensor T element $g_T \bar{\psi}_p \sigma_{\mu\nu} \psi_n$. The first EFT formulation of neutron decay (56) succeeded in reproducing the order $10^{-3}$ radiative and recoil corrections, derived before with conventional methods, see (57) and references therein. Another useful result of this EFT calculation is that the next-order corrections can be limited to be $10^{-5}$ or smaller. However, at present low-energy EFT universality cannot be tested in neutron decay because data for other processes with the same particle content like pion $\beta$ decay are not yet accurate enough. Other uses of neutron decay EFT are discussed in Sections 5 and 6.

**3.3.3. Standard Model Effective Field Theory.** In high-energy physics, no new particles beyond the SM have yet been observed, therefore searches for indirect signals have led to a revival of high-energy EFT methods called SMEFT (Standard Model Effective Field Theory) (58), initiated in 1986 (59); for recent elaborations, see (60,61). The lower scale $\Lambda_0$ of SMEFTs is not the MeV scale of $\beta$ decay but the electroweak scale $v = 246$ GeV, so the upper scale $\Lambda \gg v$ is far beyond the TeV scale and is not yet in reach experimentally. In SMEFT, Equation 5 still holds, but with a different field content in the operators $\mathcal{O}_i$, built from all known fermion and boson fields and not only from the $u, d, e, \bar{v}_e$ fields. For a list of such operators, see (55). The low-energy Wilson coefficients $\varepsilon_i$ are replaced by a larger set of SMEFT Wilson coefficients $w_j$, which are related to the $\varepsilon_i$ by appropriate matching conditions, see Equation 55 in (10).

A primary aim of contemporary particle physics is to find the EFT Wilson coefficients, be it from low- or high-energy experiments. The SMEFT approach permits comparison of limits from high-energy experiments with those from low-energy experiments, as discussed in Section 6. Experiments provide limits of the products $g_i \varepsilon_i$ or $g_j w_j$, so to determine the Wilson coefficients one needs to know the various couplings $g_i$.

## 3.4. Lattice Gauge Theory

The weak couplings $g_A$, $g_S$, $g_T$, and $g_P$ are dominated by strong-interaction effects. QCD at low energy cannot be calculated with perturbative methods because the running coupling constant of the strong interaction is no longer a small number. Today, calculations of the weak coupling constants are obtained primarily with lattice QCD theory, formulated on a large lattice of points in space and time, with periodic boundary conditions. The quark fields are placed on the lattice sites and are connected to their neighboring sites by the gluon fields. Lattice QCD relies on Monte Carlo methods and requires large computing resources. The axial coupling is calculated with a precision of ~1% to $g_A = 1.271(13)$ in Reference (62) and to ~3% in References (63-65). The numbers in parentheses give the one standard deviation error in units of the least significant digit. The results are compiled in the FLAG Review 2019 (66), which averages them to $g_A = 1.251(33)$. Scalar and tensor couplings are calculated to be $g_S = 1.02(10)$ and $g_T = 0.99(4)$.

## 4. Neutron Decay Experiments

Many clever approaches exist to measure neutron decay parameters. The format of the present review allows us to cover only a small number of these, and we concentrate on results from the past few years, whose statistics dominate earlier experiments. Only briefly listed are other experiments from the past decade and running experiments, both well covered in Reference (10), while projected experiments are covered in Reference (68). If in the following a result is given with no reference, then it is from the Particle Data Group (PDG-2020), Reference (69). For an overview of results that enter the PDG-2020 average we refer to Figure 4 in Section 5.

### 4.1. Neutron Lifetime

In the past few years, the precision of neutron lifetime measurements has considerably improved. We treat two recent lifetime measurements in more detail, and list shortly other ongoing lifetime experiments.

**4.1.1. General considerations.** Early lifetime experiments were done with neutrons in-beam (70,71), later also with UCN stored in magnetic (72,73) or material (74) bottles (traps). Contemporary neutron bottles are filled during several minutes with UCN, and then the door of the bottle is closed. For several further minutes, the spectrum is cleaned from "quasi-stable" UCNs whose kinetic energy is (usually only slightly) higher than the repulsive potential of the confining walls. To avoid quasi-stable orbits, all UCN traps have some curved or corrugated surfaces that favor unstable chaotic trajectories. After a holding time $T$, the door is reopened and the number $N(T)$ of surviving UCNs is counted. This is repeated for two or more different holding times, with $T$ ranging from a few minutes up to some fraction of an hour. If no UCNs are lost, $\tau_n$ is obtained from an exponential fit to $N(T) \propto \exp(-T/\tau_n)$, and no absolute measurement is required. For a recent review, see (75).

Lifetime experiments "in-bottle", on which we focus, may suffer from uncontrolled UCN losses. For material storage, these are due to residual inelastic wall interactions, for magnetic storage they are due to uncontrolled neutron spin flips. The energy spectrum of the stored UCN is extremely sensitive to small perturbations: Even the slow closure of the entrance port of an UCN bottle can visibly shift the lifetime result (76).

In the experiments described in the next two subsections, the UCNs are guided into the trap from below and are vertically confined by gravitation. At the end of a storage cycle, the UCNs leave the bottle through the same port to fall onto a UCN detector that is installed ~1 m below the trap.

**4.1.2. Lifetime $\tau_n$ with PNPI Gravitrap2 at ILL.** In this material bottle experiment (77), gravitational confinement limits UCN energy to 60 neV. The trap's copper walls are covered with a hydrogen-free grease and are cooled down to liquid nitrogen temperature, such that losses due to inelastic wall collisions are strongly reduced. The bottle consists of two concentric interconnected half cylinders of 2 m length, both filled with UCN, see **Figure 2a**. The inner half cylinder, the insert, has a radius of 0.6 m, the outer half cylinder, the trap, has a radius of 0.7 m. Insert and trap are independently rotatable about their common axis. Measurements are done with the trap in its lowest position, which defines the total storage volume. By changing the angle of the insert, the area of the reflecting surfaces to which the stored UCN are exposed (and with it the wall collision frequency) can be varied, while keeping the trap volume constant. This allows successive measurements with several different mean frequencies of wall collisions, whose results are extrapolated to zero frequency (i.e., to zero wall losses).

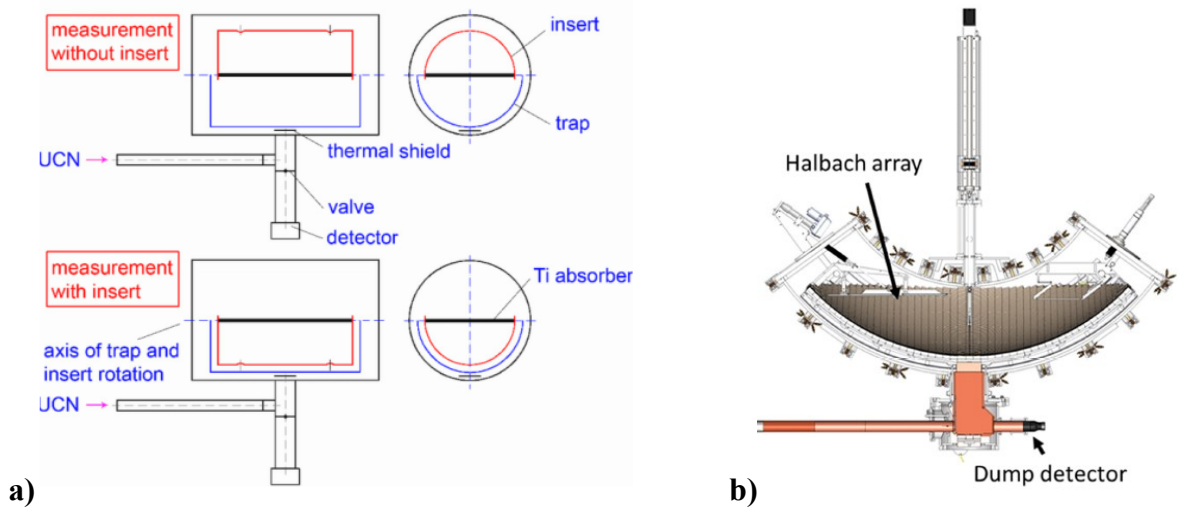

a)                      b)

**Figure 2.** Two neutron lifetime instruments. The UCN are guided into the gravitational traps from below. At the end of a counting cycle, they leave the bottle downward into the UCN detector. **a)** Cross sections of the PNPI bottle at ILL. Panel with permission from Reference (77); copyright 2017 American Physical Society. **b)** Cross section of the UCN$\tau$ experiment at LANL. Panel with permission from Reference (81); copyright C.L. Morris et al./CC BY 4.0.

At the end of each measurement with a fixed mean collision frequency, the energy spectrum of the surviving UCN, whose height distribution follows the barometric formula, is determined by pouring the UCN stepwise out of the trap onto their detector. This "decanting" is done by first putting trap and insert into their lowest position, and then rotating (tilting) them together through different angles, each time counting the outflowing UCNs. The UCN energy intervals corresponding to the successive tilt angles are 0-40.1 neV, 40.1-50.5 neV, and 50.5-56.6 neV. The total process is simulated by Monte Carlo, which successfully reproduces the UCN rates over more than four orders of magnitude (78). The longest measured storage time is only 2.5 s below the extrapolated result $\tau_n = 881.5(0.7)(0.6)$ s (the first error is statistical, the second systematic). Previous PNPI lifetime results (79) were far below the PDG averages of the time and were met with disbelief (80), but were later confirmed by the experiments described below.

**4.1.3. Lifetime $\tau_n$ with the UCNτ trap at LANL.** In this magnetic bottle experiment (82), gravitational confinement limits UCN energy to 38 neV. The trap is a shallow bowl of 600 liter volume formed by an array of permanent magnets, with a 10 mT holding field superimposed, see **Figure 2b**. The array is asymmetrically shaped to avoid quasi-stable UCN orbits. UCN losses are excluded based on field maps measured near the magnetic surface, see (83) and references therein.

A special feature of the apparatus is that the surviving UCN are counted in situ with a small UCN detector movable vertically inside the storage volume, which allows energy resolved monitoring of UCN phase-space evolution. This monitor is a $^{10}$B-coated scintillator, which measures either the $\alpha$ or the $^7$Li, emitted back to back after neutron capture. The scintillator is coupled via two flexible light guides to photomultipliers, one at each end. A detector of the same type is used for active cleaning of UCN on quasi-stable orbits. The result of the blinded measurement is $\tau_n = 877.7(0.7)(^{+0.4}_{-0.2})$ s. Corrections to the lifetime are smaller than the error. The largest systematic error of one quarter second is due to microphonic heating.

**4.1.4. Other lifetime experiments.**
– The PNPI-magnetic trap at ILL (84) is built from an upright cylinder of permanent magnets. A special feature of the experiment is that the UCNs are transported into the decay volume from above via an UCN "lift". This avoids changes to the UCN spectrum that occur if, instead, the magnetic shutter at the trap's bottom is used for filling. UCNs that suffer accidental spin-flips leave the trap through this magnetic shutter and are counted separately. The result is $\tau_n = 878.3(1.6)(1.0)$ s.

– The Kurtchatov material-wall trap at ILL (76) monitors upscattered UCNs in neutron detectors installed outside the bottle. As in the gravitational PNPI trap, the effective surface area of the trap, and with it the collision frequency, is enlarged at constant volume, here by inserting large numbers of oil-covered copper strips. The result is $\tau_n = 880.2(1.2)$ s.

– The in-beam experiment at NIST (85) relies on absolute measurements of the average number of cold neutrons in the decay volume, of the number of decay protons, and of the proton and neutron detector efficiencies, from which $\tau_n = 887.7(1.2)(1.9)$ s is derived, about four standard deviations (sigma or $\sigma$) above the PDG-2020 average.

PDG-2020 averages the four latest bottle measurements described above plus two material bottle results (86,87), the latter reevaluating the first 1986 UCN lifetime measurement (74), and derives a lifetime average $\tau_n = 879.4(0.6)$ s where the error is increased by a scale factor $S = 1.6$ to account for deviate results.

**4.1.5. Upcoming lifetime experiments.** Future UCN in-bottle experiments are:

– PENeLOPE at FRM II (88),

– HOPE at ILL (89), and

– τSPECT at TRIGA-Mainz (90), all with magnetic storage.

New in-beam experiments are prepared at J-PARC (91), with a first result $\tau_n = 898(10)(^{+15}_{-18})$ s, and at NIST (92).

## 4.2. Neutron Decay Correlations

In recent years, the precision of neutron correlation measurements has strongly improved as well. We treat a few recent correlation measurements in more detail, and list shortly other ongoing correlation experiments.

**4.2.1. General considerations.** Within the SM, all correlations depend only on $\lambda = g_A/g_V$, as in Equation 4. The parity-violating $\beta$ asymmetry $A$ and the $e$-$\bar{\nu}_e$ correlation coefficient $a$ measure the deviation of $|\lambda|$ from unity with (about equal) high sensitivity for $\lambda$. We limit our discussion to two recent $\beta$ asymmetry $A$ and one $\beta$-$\nu$ correlation $a$ measurement, and quote first neutron limits on the possible presence of a Fierz interference term $b$.

In these experiments, the charged decay particles are spiraling adiabatically about a magnetic guide field, which connects the neutron decay volume with the detector(s). In the two asymmetry experiments, PERKEO III and UCNA, two energy-sensitive electron detectors are positioned symmetrically at both ends of the apparatus. These are fast plastic scintillators, which have low sensitivity to gamma ray background and offer fast timing, as required for electron backscatter detection in the opposite detector. The photons of the scintillators are read out by photomultiplier tubes; an alternate scheme for photon readout with higher energy resolution is proposed in Reference (93). A time-of-flight method for detector characterization was developed in Reference (94). In this context, the detection of synchrotron radiation from gyrating $\beta$ particles may also become interesting, see (95) and references therein.

Magnetic transport has the advantage that for a decay event that occurs at an arbitrary position $x$ of the decay volume, the solid angle of initial emission along or opposite to the field direction is always exactly $2\pi$, which makes such asymmetry measurements independent of the precise local orientation $\boldsymbol{B}(\boldsymbol{x})$ of the field and of the detector position. In both $\beta$ asymmetry experiments, the guide field decreases toward the detector positions to exclude glancing incidence of the electrons on the detector, by means of the inverse magnetic mirror effect. The field decrease also avoids local field minima within the decay volume, in which the electrons could be temporarily trapped and could be lost or assigned to the wrong hemisphere. The spatial distribution of charged decay particles on the detector surface after magnetic transport was calculated and tested with electrons in Reference (96). Both experiments on $A$ use a blinding scheme for data evaluation to eliminate potential bias.

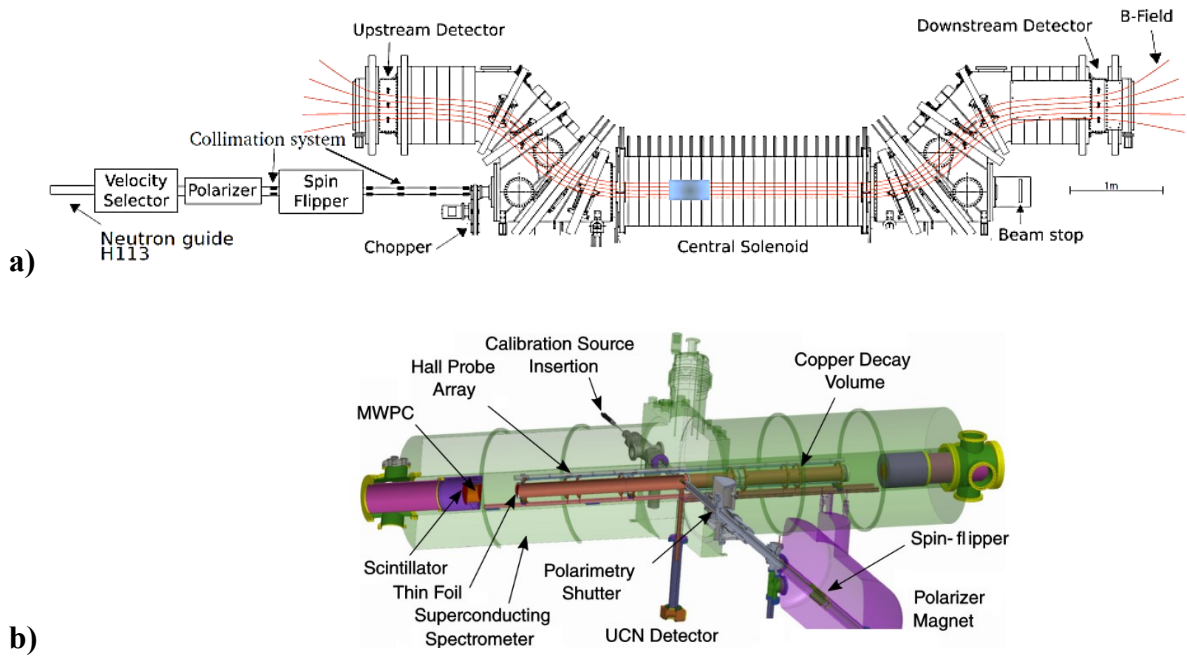

**Figure 3.** Two neutron $\beta$ asymmetry instruments. **a)** PERKEO III: A pulse of cold neutrons (shown symbolically in light blue) enters the decay volume from the left. Note the 1 m-scale on the right. Panel adapted with permission from Reference (97); copyright 2019 American Physical Society. **b)** UCNA: The UCN enter the 3-m-long decay volume (in brown) from the forefront. Panel adapted with permission from Reference (99); copyright 2018 American Physical Society.

**4.2.2. β asymmetry *A* with PERKEO at ILL.** The PERKEO III instrument (97), **Figure 3a**, uses a monochromatized beam from ILL's ballistic supermirror neutron guide, polarized to 99.10(6)%, and pulsed by a mechanical disc chopper with a duty cycle of 1:14. A high-density "cloud" of cold neutrons is moving freely along the beam axis through the instrument, without meeting any material obstacle, and is finally absorbed in a beam stop. Neutron decay is observed only when the neutron pulse is fully contained within the fiducial decay volume, whose length of ~2 m can still be optimized during later data analysis. A field of $B = 0.15$ T from normally conducting coils projects the decay electrons without any edge effects onto their detectors. From the peak electron rate $n_\beta \gtrsim 1000$ s$^{-1}$ it follows that the number of cold polarized neutrons in each pulse is $N_n = n_\beta \tau_n \gtrsim 10^6$. After the neutron pulse is completely absorbed by the beam stop, an additional time window allows measuring the background under the same conditions as the signal, i.e., with the neutron beam chopper closed. This reduces uncertainties related to the background measurements to the $10^{-4}$ level. The result is $A = -0.11985(17)(12)$.

**4.2.3. β asymmetry *A* with UCNA at LANL.** The UCNA experiment (99), **Figure 3b**, uses a dedicated UCN source driven by a proton accelerator, polarized to 99.79(15)%. Typically, 4000 UCNs are stored in a horizontal 3-m long cylindrical bottle with material walls and a superconducting guiding field $B = 1$ T. The bottle is closed by two thin windows that have a combined thickness of 300 nm. To suppress background, low-pressure wire chamber detectors closed by two 6-μm mylar windows are added, in coincidence with the subsequent electron detectors. The additional electron interaction is addressed with Monte Carlo simulations. A continuous electron rate of $n_\beta = 25$ s$^{-1}$ is obtained at a very low background of 0.025 s$^{-1}$. The result is $A = -0.12015(34)(63)$.

**4.2.4. Electron-antineutrino correlation *a* with *a*SPECT at ILL.** The correlation *a* is inferred from the recoil spectrum of decay protons from a beam of unpolarized cold neutrons (100). A variable electrostatic potential superimposed on the magnetic guide field serves as a barrier to block the lower part of the energy spectrum from reaching the proton detector. An integral spectrum at different blocking potentials is recorded, similar as in the KATRIN experiment on the neutrino mass (101). To determine the impact of systematic effects, measurements were made for seven different configurations of the *a*SPECT apparatus, and key properties of the instrument were deliberately changed from their optimum settings. A simultaneous fit to all these measurements was done, including a full simulation of the apparatus. As a final result of the global fit, $a = -0.10430(84)$ with $S = 1.2$ was obtained, which results in a relative uncertainty of Δ$a/a$ = 0.8%, five times more precise than the best previous experiment. All values from the seven individual configurations, including systematic corrections as well as their weighted mean, are statistically in agreement with this result. Taking the average of the early results (102, 103), of the above aSPECT result (100), and of the new aCORN result (104) $a = -0.10782(181)$ (blinded analysis) gives the new global average $a = -0.10486(75)$.

**4.2.5. Fierz interference $b_n$.** The Fierz term $b$, which vanishes in the SM, multiplies $\beta$ decay spectra by a factor $1 + b\,mc^2/(E + mc^2)$, which decreases with increasing electron energy $E$. For limits on $b$ from various nuclear sources, see Reference (105). A search in the unpolarized neutron's $\beta$ spectrum by the UCNA collaboration (106) gave $b_n = 0.067(5)(^{90}_{-61})$. A search in the polarized $\beta$ asymmetry spectrum, where $b_n$ enters as $A(E) = A_0(E)/[1 + b_n\,mc^2/(E + mc^2)]$, gave $b_n = 0.017(20)(3)$ from PERKEO III (98), and $b_n = 0.066(41)(24)$ from UCNA (107). The electron neutrino coefficient $a$ is not sensitive to $b_n$, except for secondary data extraction effects (5).

**4.2.6. Correlation coefficients $B$, $C$, $D$, $R$, $N$:**

– The neutrino asymmetry $B$ must be detected via e-p coincidence. The PDG-2020 average is $B = 0.9807(30)$, and there have been no new measurements in more than a decade. The individual results for $B$, see (108-110), are consistent with each other, but there is a $2\sigma$ tension with the expected value $B_{SM} = 0.98710(8)$ obtained when our average for $\lambda$ is inserted into Equation 4.

– The proton asymmetry $C = x_C(A + B)$, with $x_C$ = 0.27484, is not an independent observable. The only value $C = 0.2377(26)$ from PERKEO II, see (111), is measured in e-p coincidence in two detectors, with each detectors being sensitive to both protons and electrons, and is based on the same data set as for $B$ in Reference (110).

– The non-SM $T$-odd (parity even) triple correlation coefficient $D$, which depends on Im($\lambda$), was measured to $D = -2.8(6.4)(3.0) \times 10^{-4}$ at ILL (112) and $D = -0.94(1.89)(0.97) \times 10^{-4}$ at NIST (113), using separate $p$ and $e$ detectors arranged under different angles around a cold neutron beam. The PDG-2020 average is $D = -1.2(2.0) \times 10^{-4}$.

– The non-SM $T$-odd and $P$-odd triple correlation coefficient $R$ and the SM-allowed coefficient $N$ both involve the electron spin $\sigma_e$. The nTRV collaboration at PSI (114) used an arrangement of $p$ and $e$ detectors similar as for the $D$ coefficient and succeeded for the first time in installing electron spin analysis by Mott scattering on thin lead foils, with the results $R = 0.004(12)(5)$ and $N = 0.67(11)(5)$.

**4.2.7. Rare Standard Model decay channels.** Radiative neutron decay $n \to p^+ e^- \bar{\nu}_e \gamma$ is accompanied by the emission of a bremsstrahlung photon. Its branching ratio of ~1% was precisely measured with the NIST in-beam lifetime apparatus (115). The photon's polarized spectrum was calculated by EFT (116).

Another rare neutron decay channel is bound $\beta$ decay $n \to \text{H} + \bar{\nu}_e$, where the emitted electron ends up in a bound S-state of hydrogen H with a branching ratio of $4 \times 10^{-6}$. For an experiment in preparation, see Reference (117).

**4.2.8. Upcoming correlation experiments.** The following instruments are running or are being planned.

– Nab at SNS (118,119) will use the time of flight of electrons and protons to determine the electron-antineutrino coefficient $a$ and the Fierz term $b$ from cuts to the Dalitz plot.

– The proton asymmetry $C$ can be measured without coincidence at high count rates by detection of protons only. Recent results from PERKEO III are still blinded.

– PERC is a beam station that will deliver not neutrons but an intense beam of neutron decay products, extracted from inside a neutron guide, for measurements of decay correlations by interested experimental groups (120,121).

– NoMoS is an $R \times B$ drift momentum spectrometer to be installed at PERC (122).

– ANNI will be a cold beam station at the ESS dedicated to neutron-particle physics (123). By exploiting the time-structure of ESS, a successor to PERC could provide more than an order of magnitude improved statistics.

– BRAND generalizes the nTRV concept of electron polarization measurement, see Section 4.2.6, and proposes to measure simultaneously 11 correlation coefficients (124) in one single run.

## 5. Applications of $\beta$ Decay Data

### 5.1. Applications Other than Particle Physics

The rates of all weak processes that involve both leptons and quarks must be calculated from measured neutron data. We first mention some applications outside of particle physics. In big bang nucleosynthesis, the neutron lifetime enters twice: in the neutrino cross section $\sigma_\nu \propto 1/\tau_n$, which determines at what time the early Universe falls out of equilibrium (~1 s); and thereafter in the decaying number of neutrons $\propto \exp(-t/\tau_n)$ available for element production. The neutron lifetime contributes the largest error in the calculation of the primordial $^4$He mass fraction, see table II in (125). However, this theory error is markedly smaller than the observational error of stellar $^4$He abundance, see Review 24 "Big Bang Nucleosynthsesis" in PDG-2020. Hence, for the time being, the experimental lifetime $\tau_n$ is precise enough for this application. The same holds for solar and stellar temperatures, which depend on neutron weak interaction data, and also for the efficiencies of solar neutrino detectors based on inverse neutron decay. Therefore, we refer the reader to our earlier review on these topics (3), and treat in the following applications in particle physics only.

## 5.2. Results within the Standard Model

In this Section, we list results on SM weak interaction parameters derived from neutron, nuclear, and pion $\beta$ decay experiments.

### 5.2.1. Results from neutron $\beta$ decay. Figure 4a shows the PDG-2020 data for the neutron lifetime $\tau_n$ and $\lambda$ (updated), whose averages and scale factors are

$$\tau_n = 879.4(0.6), \ S = 1.6, \ \lambda = -1.2754(11), \ S = 2.2 \qquad 6.$$

The most recent results for $\lambda$ from $A$ are in excellent agreement with each other, the last two using blind analyses. However, there is some tension between with $a$ from aSPECT These values for $\tau_n$ and $\lambda$ can be used in Equations 3 and 4 to derive $|V_{ud}|_n = 0.97377(33)_\tau(70)_\lambda(11)_{RC}$, where RC stands for radiative corrections, or $|V_{ud}|_n = 0.97377(78)$.

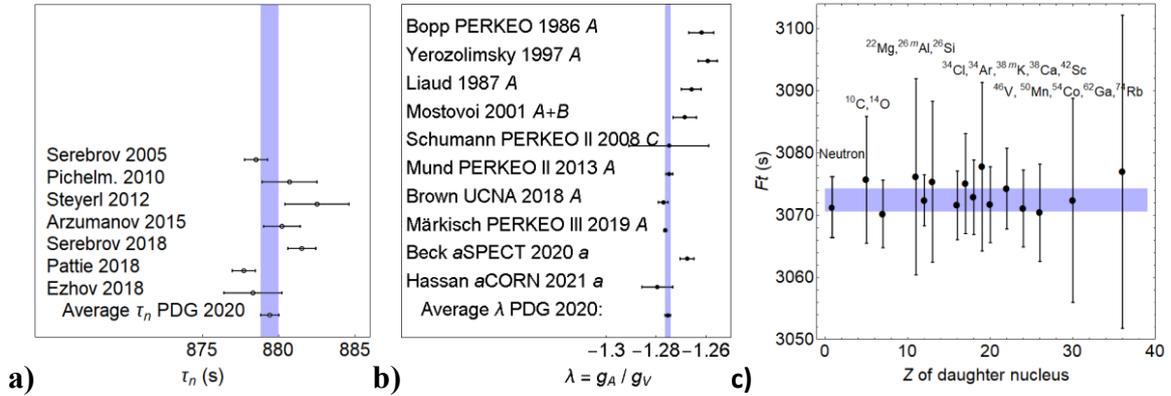

**Figure 4.** Results of neutron decay experiments. **a)** $\tau_n$ and **b)** $\lambda = g_A/g_V$. The blue vertical bands indicate the world averages of PDG-2020. The corresponding references are cited in PDG-2020 under the names indicated in panels a) and b). (In the case of $\lambda$, the PDG-2020 data are updated with $a$ from References (100,104). **c)** $Ft$ values for superallowed $\beta$ decays, which include recent changes in nuclear theory, and their average, Equation 7, indicated by the blue horizontal band. The neutron's $Ft_n^V$, Equation 8, is added at $Z = 1$.

In Equation 3 on $\tau_n$, the radiative correction $\delta'_R$ and $\Delta_R$ are needed. The so-called outer radiative correction $\delta'_R$ depends only on the electron energy and charge number $Z$ of the daughter nucleus. For the neutron, the phase factor is $f = 1.6887(2)$, and $\delta'_R = 1.014902(2)$ from Reference (126), hence $f(1 + \delta'_R) = 1.7139(2)$. $\Delta_R$ is the transition independent part of the radiative corrections, which is the same for all nuclei, including the neutron. $\Delta_R$ has recently been reevaluated using dispersion relation techniques (127,128), leading to a reduced error and a $3\sigma$ shift from the previous value to $\Delta_R = 0.02477(24)$.

$\beta$ transitions are often characterized by their $ft$ values, with a measured half-life $t$ and a calculated phase factor $f$. When the transition specific corrections are applied to the $ft$ values, one obtains the so-called $Ft$-values. To permit comparison with nuclear $Ft$-values below, we write, with a spin factor ½, the corresponding neutron value $Ft_n = ½ \ln 2 f \tau_n (1 + \delta'_R) = 522.35(36)$ s. As $\Delta_R$ is universal, it does not enter the definition of $Ft$.

**5.2.2. Results from nuclear $\beta$ transitions.** Nuclear superallowed $0^+ \to 0^+$ $\beta$ transitions have $Ft_{0-0} = ft_{0-0}(1 + \delta'_R)(1 + \delta_{NS} - \delta_C)$ with additional transition-dependent nuclear structure corrections $\delta_{NS}$ and isospin corrections $\delta_C$, which are of similar size as $\delta'_R$, (typically ~1.5%) but less well known. Hardy and Towner (129) found for the weighted average from 15 isotopes $\overline{Ft}_{0-0} = 3072.24(0.57)_{\text{stat}}(0.36)_{\delta'_R}(1.73)_{\delta_{NS}}$ s, or

$$\overline{Ft}_{0-0} = 3072.24(1.85) \text{ s.} \qquad 7.$$

Using CVC, the neutron and the nuclear $Ft$ values should be the same, but this holds only for the vector part of the neutron $Ft$ value, whose branching ratio is $1/(1 + 3\lambda^2) = 17\%$, or

$$Ft_n^V = (1 + 3\lambda^2) Ft_n = 3071.4(4.9) \text{ s}, \qquad 8.$$

in good agreement with $\overline{Ft}_{0-0}$.

We display $Ft(Z)$ in **Figure 4c**, starting at $Z = 1$. For the $Ft_{0-0}$ values with $Z > 1$, we do not use figure 3 from Reference (129), which shows the statistical errors but not the theory errors from newly discovered distortions of the outgoing electron spectrum by nuclear polarizabilities, accompanied by a global nuclear polarizability correction (130,131). In Figure 4c we include these theory errors, which were obtained from the bar graph of figure 4 of Reference (129).

The sizes of nuclear structure corrections applied to the $ft_{0-0}(Z)$ data are 10 to 20 times the theory errors. In view of this fact, the excellent consistency of the final $Ft_{0-0}$ results, is a triumph of nuclear theory. We see that the neutron's $Ft_n^V$ competes well with the nuclear $Ft_{0-0}(Z)$, and as discussed below, both are needed and complement each other.

The CKM matrix element $V_{ud}$ is obtained from $Ft_{0-0}(Z)$, using the relation $|V_{ud}|^2_{0-0} = = 2984.4(1.1) \text{ s}/[Ft_{0-0}(1 + \Delta_R)]$ from Reference (129). At present, this nuclear value is

more precise than the neutron value by a factor 2.5, and an order of magnitude better than the value from the rare ($1.0\times10^{-8}$) pion decay (132). Altogether,

$$|V_{ud}|_{0-0} = 0.97373(31), |V_{ud}|_n = 0.97377(78), |V_{ud}|_\pi = 0.97390(290). \qquad 9.$$

Even better limits are obtained from a global fit (133) with up to 14 parameters using EFT, based on 6 neutron and more than 30 nuclear observables. This global fit, with a reduced chi-square $\chi^2 = 0.8$, sharpens the neutron lifetime such that its additional dependence on $\lambda$ leads to a 2.4 times better value for $\lambda$:

$$\lambda_{\text{total}} = -1.27529(45), \text{ and } |V_{ud}|_{\text{total}} = 0.97370(25). \qquad 10.$$

## 5.3. Limits beyond the Standard Model

With EFT, large quantities of weak interaction data can be combined within a single fit, see (10,133,134), so why should we bother to evaluate neutron and other data separately? The answer is that each data source has its specific strengths and weaknesses, both from theory and experiment, and it is good practice to have a separate look on them. Besides, overly tight guidance by perspective outlooks may block serendipitous discoveries.

In the following subsections, limits on exotic processes and the corresponding Wilson coefficients are mostly taken from the reviews in References (9) and (10), some of which are updated in Reference (133).

**5.3.1. Cabibbo-Kobayashi-Maskawa unitarity.** The first row of the CKM matrix involving up quarks $u$ is the most sensitive for finding deviations from CKM unitarity, with

$$\Delta^u_{\text{CKM}} \equiv |V_{ud}|^2 + |V_{us}|^2 + |V_{ub}|^2 - 1 = -0.0012(7), \text{ or } = -0.0021(7), \qquad 11.$$

depending on whether one derives $V_{us}$ from $K \to \mu\bar{\nu}_\mu$ or from $K \to \pi e\bar{\nu}_e$ decays, see table 1 in (135). This is two to three standard deviations away from the SM expectation. Limits on $\Delta^u_{\text{CKM}}$ not from the first row but from the first column are about three times, and those on $\Delta^s_{\text{CKM}}$ from the row or column of the second family are about 40 times weaker; see review 12, "CKM Quark Mixing Matrix" in PDG-2020. A fourth particle family $N_\nu = 4$ as a source of CKM unitarity violation seems unlikely in view of the $e^+e^-$ collider limit $N_\nu = 2.984(8)$, unless the fourth neutrino has a huge mass.

Deviations from CKM unitarity $\Delta^u_{\text{CKM}} \neq 0$ can be induced by the Wilson coefficients $\varepsilon_L$ and $\varepsilon_R$ or by an exotic shift of the Fermi constant $G_F$ obtained from muon decay, which would lead to a $V_{ud}$ shift of size $\Delta|V_{ud}|^2 = |V_{ud}|^2 (\varepsilon_L + \varepsilon_R - \delta G_F/G_F)$ and to a corresponding $V_{us}$ shift of size $\Delta|V_{us}|^2$. These shifts are strongly constrained by neutron and nuclear $\beta$ decay measurements to $\Delta|V_{ud}|^2 + \Delta|V_{us}|^2 = -0.0001(14)$, from Reference (10) (based on PDG-2018 data).

**5.3.2. Dark neutron decay.** Recently, the 4$\sigma$ difference between in-beam and in-bottle neutron lifetime results provoked a large number of investigations, initiated by Fornal and Grinstein (136), asking whether this "neutron anomaly" might be due to additional neutron decays into exotic dark particles. Dark decays would shorten the measured in-bottle lifetime but not the measured in-beam lifetime. Using the new neutron data, such dark decays can be excluded as follows (137). Under CVC we expect that $Ft_n^V \equiv \overline{Ft}_{0-0}$, or, from Equation 8, that ½ ln2 $f\tau_n(1+3\lambda^2)(1+\delta_R') = \overline{Ft}_{0-0}$, from which the neutron lifetime is calculated as

$$\tau_n^\lambda = \frac{2}{\ln 2} \frac{\overline{Ft}_{0-0}}{f(1+\delta_R')(1+3\lambda^2)} = \frac{5172.3(3.1) \text{ s}}{1+3\lambda^2}. \qquad 12.$$

An advantage of this SM link between neutron lifetime and $\lambda$ is that it shows explicitly the independence of $\Delta_R$ and the uncertainties related with it, a fact not visible in the "SM master formula" $|V_{ud}|^2 \tau_n(1+3\lambda^2)(1+\Delta_R) = 4900.1(1.1)$ s, updated from Reference (138). With the new $\lambda$ of Equation 6, we obtain $\tau_n^\lambda = 879.65(62)$ s. This value agrees with the bottle-lifetime average $\tau_n = 879.4(0.6)$ s, which would be shifted if dark decays existed, but not with the in-beam lifetime 888.0 (2.0) s, which is insensitive to dark decays. This excludes dark neutron decays as an explanation of the neutron anomaly at a level of 4$\sigma$.

**5.3.3. Potential constraints from axial coupling $g_A$.** To isolate an exotic shift of $g_A$ like $g_A(1-2\varepsilon_R)$ induced by right-handed quark currents, one needs an SM prediction for $g_A$. The sensitivity to such a shift is limited to ~1% by the error of $g_A$ from lattice QCD, see Section 3.4, not by the ~$10^{-3}$ error of the measured $g_A$.

**5.3.4. Limits on scalar S and tensor T coupling.** S and T couplings $g_S \varepsilon_S$ and $g_T \varepsilon_T$ enter $\beta$ decay linearly via the Fierz interference term $b$, and we know from lattice QCD, Section 3.4, that the couplings $g_S$ and $g_T$ are close to one. A global fit to the 15 pure Fermi $Ft_{0-0}$ values gives a precise limit $b_F = -0.0000(20)$ from (129). This Fierz term depends on the scalar Wilson coefficient $\varepsilon_S$ as $b_F = [1-(\alpha Z)^2]^{1/2} g_S \varepsilon_S$ (with $\alpha = 1/137$), and its limit translates into $\varepsilon_S \approx g_S \varepsilon_S = -0.0000(10)$.

The neutron Fierz term $b_n$ depends both on scalar and tensor Wilson coefficients, $b_n = (1-\alpha^2)^{1/2}(g_S \varepsilon_S + 12\lambda g_T \varepsilon_T)/(1+3\lambda^2)$, which can be disentangled by using the above nuclear limit on $g_S \varepsilon_S$. The results from Section 4.2.5 on $b_n$ from the neutron $\beta$ asymmetry spectrum $A(E)$ give $\varepsilon_T = -0.007(8)$ from PERKEO III and $\varepsilon_T = -0.025(18)$ from UCNA, which we combine to $\varepsilon_T = -0.009(7)$. As pointed out previously (3), much better limits on the Fierz terms are expected from neutron decay in a joint fit with $B(E)$ and $C(E)$ data, whose strongly elongated $\chi^2$-contours are nearly orthogonal to those of $A(E)$ and $a(E)$.

A very tight limit $\varepsilon_T = -0.00028(53)$ is obtained from the ratio of the neutron $\beta$ asymmetry to that of $^{19}$Ne (139,140). Even better limits come from the global fit (133) mentioned at the end of Section 5.2.2, to which we add the limit on pseudoscalar interactions from pion decay (10),

$$\varepsilon_S = -0.0000(10), \varepsilon_T = 0.0001(3), \varepsilon_P = 0.000004(13), \qquad 13.$$

and conclude these limits do not leave much room for deviations from the V–A standard model. Note that for $\varepsilon_T$, we stick to the definition of Reference (10), which differs from that of Reference (133).

**5.3.5. Limits on time reversal violation.** *CP* violation, and with it *T* violation, as observed for Kaons and *B*-mesons, is accommodated in the SM via a complex phase factor $e^{i\delta}$ in the CKM matrix, with $\delta = 1.20(5)$ (from review 12 in PDG-2020). These SM-effects are too small to be observable in $\beta$ decay, and the same is true for a *T*-violating EDM. In both cases, a nonzero effect would therefore be a unique sign of New Physics beyond the SM.

The *T*-violating neutron triple correlation coefficient $D = -.00012(20)$ from Section 4.2.6 is accompanied by finite-state interactions $\sim 10^{-5}$. For EFT calculations on *D*, see (141). T invariance requires coupling constants to be real, with relative phases being either 0 or 180°. With $\sin\phi_{AV} = D(1 + 3\lambda^2)/(2|\lambda|)$, this *D* value corresponds to a phase $\phi_{AV} = 180.017(26)°$ of $g_A$ relative to $g_V$, which translates into limits on Wilson coefficients $|\text{Im}(\varepsilon_{L,R})| < 0.0004$ at 90% confidence limit (CL).

The neutron triple correlation coefficient $R = 0.004(12)(5)$ from Section 4.2.6, with final state interactions $\sim 4 \times 10^{-4}$, leads to limits on a *T*-violating phase $\phi_{TS}$ of $g_S$ relative to $g_T$. Its S and T Wilson coefficients can be disentangled by using $R = 0.009(22)$ from $^8$Li decay, which gives a purely tensor limit $\text{Im}(\varepsilon_T) < 0.0015$, leading to $|\text{Im}(\varepsilon_S)| < 0.06$ for the neutron. Under rather general assumptions it is possible to translate bounds on *D* and *R* into bounds on the neutron EDM with much stronger $\sim 10^{-6}$ limits from the EDM, see table VII of Reference (9).

**5.3.6. Searches for right-handed quark currents.** A decade ago, for the simplest case of left-right symmetry "manifest" in the Hamiltonian (142), exclusion plots for right-handed $W_R$ mass $m_R$ vs. left-right mixing angle $\zeta$ had been derived from neutron $\beta$ decay data (3), resulting in a mass limit $m_R <$ 250 GeV and $-0.23 < \zeta < 0.06$ at 95% CL. In addition, from the limit on CKM unitarity violation follows $|\zeta| \lesssim 0.003$, see figure 3 in (143). These limits can be translated into Wilson coefficients, the mixing angle $\zeta$ corresponding to $-\varepsilon_R$, and the right-to-left mass ratio squared corresponding to $\tilde{\varepsilon}_R$ (54). A nonzero $\zeta$ would change $g_A$ by a factor $1 - 2\varepsilon_R$, and $V_{ud}$ by 1 − $\varepsilon_R$ (10). The new neutron data do not change these limits significantly, and therefore we do not reopen this topic.

**5.3.7. Tests of Lorentz invariance.** Attempts to unify the SM with general relativity often lead to a spontaneous breaking of Lorentz invariance, a process parametrized in Reference (144). Such a violation can be observed as a sidereal variation of many possible observables; for a recent list of experiments, see (145). Searches for daily variations of neutron and nuclear $\beta$ decay observables (9) have obtained sub-percent limits on such Lorentz violation (146).

In particle physics, the number of new models is almost unlimited, and so is the number of constraints that may be derived from neutron decay, such as lepton flavor universality (147), so we end our discussion at this point.

# 6. Comparison with High-Energy Limits

## 6.1. The Standard Model Effective Field Theory Approach

SMEFTs, with $\Lambda \gg \Lambda_0 = 246$ GeV as discussed in Section 3.3.3, are built on a large number of field configurations with corresponding unknown Wilson coefficients $w_i$, see table 100 in (134). However, being universal, SMEFT studies can make global fits to even larger numbers of experimental data, ranging from atomic *CP*-violating effects (148) through nuclear and particle $\beta$ decays all the way to the large trove of high-energy data. We have stated that low-energy EFTs are universal and model independent, but it must be kept in mind that SMEFTs require several assumptions:

– The energy gap between $\Lambda$ and $\Lambda_0$ must be large, as it is guaranteed in low-energy EFTs with $\Lambda \sim m_W \gg E \sim \Lambda_0$.

– No exotic particles must exist with masses below $\Lambda_0$ (an exception is right-handed neutrinos).

– Exotic particles must be weakly coupled so that electroweak symmetry is linearly realized;

– At high energies, dimension 8 operators with their even larger number of field configurations may not be negligible. Furthermore, the LHC limits quoted below assume that only one operator at a time is present ("sole source" vs. "global analysis").

EFT techniques permit comparison of limits for processes with the same Feynman diagram. An example are $pp$ collisions with $d\bar{u} \to e\bar{\nu}_e$, with the neutrino seen as missing transverse energy, which have the same diagrammatic representation as neutron decay $d \to ue\bar{\nu}_e$. When comparing low- and high-energy Wilson coefficients $\varepsilon_i$ and $w_j$, the running of the coupling constants with energy must be taken into account, and the high-energy coefficients $w_j$ must be translated to low-energy coefficients $\varepsilon_i$ and $\tilde{\varepsilon}_i$, or vice versa, via appropriate matching conditions, see Section 3.3.3.

## 6.2. Limits on Non-Standard Model Wilson Coefficients

**Table 5. Comparison of Wilson coefficients on beyond-SM physics**

**a) From neutron $\beta$ decay**

| Type of interaction | Wilson coefficient | Type of measurement | Ref. |
|---|---|---|---|
| Tensor | $|\varepsilon_T| < 0.0003(5)$ | $A$(neutron)/$A$($^{19}$Ne), $A$=$\beta$-asym. | (139) |
| Tensor | $\varepsilon_T = -0.0001(3)$ | All neutron and all nuclear data | (133) |
| Right-handed quark | $\varepsilon_R = 0.00(2)$ | Neutron and nuclear $Ft$ | (10) |
| Left-, Right-hd $q$, $\cancel{T}$ | $|\text{Im}(\varepsilon_{L,R})| < 0.0004$ | $D$(neutron) | 5.3.5 |
| Left-, Right-hd $q$, $\cancel{T}$ | $|\text{Im}(\varepsilon_{L,R})| < 0.000006, <0004$ | EDM(neutron) | (9) |
| Scalar, Tensor, $\cancel{T}$ | $|\text{Im}(\varepsilon_{S,T})| < 0.06, <0.0015$ | $R$(neutron), with $\varepsilon_T$ from $R$($^8$Li) | (114) |

**b) From nuclear and pion $\beta$ decay, neutrino mass**

| Scalar | $\varepsilon_S = 0.0000(10)$ | Superallowed $Ft$ | (129) |
|---|---|---|---|
| Tensor | $-0.0011 < \varepsilon_T < 0.0014$ | Radiative pion $\pi \to e\nu\gamma$ | (5) |
| Pseudoscalar | $\varepsilon_P = 0.000004(13)$ | Ratio $(\pi \to e\nu)/(\pi \to \mu\nu)$ | (10) |
| Pseudoscalar, $\cancel{T}$ | $\text{Im}(\varepsilon_P), |\varepsilon_P| < 0.0003$ | Ratio $(\pi \to e\nu)/(\pi \to \mu\nu)$ | (5) |
| L, T: Right-hd. neutrino | $|\tilde{\varepsilon}_L| \lesssim 0.01, |\tilde{\varepsilon}_T| \lesssim 0.0005$ | Neutrino mass bound | (9) |

**c) From proton collisions at LHC**

| Scalar, Pseudoscalar | $|\varepsilon_S|, |\varepsilon_P| < 0.006$ | CMS $pp \to e\nu$ transvs. energy | (149) |
|---|---|---|---|
| Scalar, Pseudo, Tensor | $|\varepsilon_S|, |\varepsilon_P| < 0.005, |\varepsilon_T| < 0.0006$ | ATLAS $pp \to e^+e^-$ energy tail | (150) |
| S, P: Right-hd. neutrino | $|\tilde{\varepsilon}_S|, |\tilde{\varepsilon}_P| < 0.006$ | ATLAS $pp \to e^+e^-$ energy tail | (150) |
| Right-hd. $q$, Right-hd. $\nu$ | $|\tilde{\varepsilon}_R| < 0.002$ | ATLAS $pp \to e^+e^-$ energy tail | (150) |

The subscript L and R stand for left- and right-handed quark currents, the tilde stands for right-handed neutrinos. $A$, $D$, and $R$, referred to in the third column, are correlation coefficients. $\cancel{T}$ indicates time reversal violation. Errors in equalities are given at $1\sigma$ or 68% CL; errors in inequalities are given at 90% CL or $1.64\sigma$.

**Table 5** lists the best results on the EFT Wilson coefficients $\varepsilon_i$ and $\tilde{\varepsilon}_i$, both from low-energy and high-energy experiments (deduced at the renormalization scale $\mu = 2$ GeV in the minimal subtraction scheme $\overline{MS}$). We rely on Refs. (5) and (10) for the translation of high-energy data into limits on $\varepsilon_i$ and $\tilde{\varepsilon}_i$.

The LHC limits given in Table 5 are obtained at a proton collision energy of 8 TeV and will improve with the LHC results taken at 13 TeV. In the new ATLAS release (151), the integrated luminosity is increased from 20 fb$^{-1}$ to 139 fb$^{-1}$, and also the limits from low-energy experiments are continuously improving. Limits on a Wilson coefficient do not simply translate into limits on an energy scale $\Lambda$, which latter must be based on a specific model. Dimensional reasoning suggests a scaling $\varepsilon_i, \tilde{\varepsilon}_i \sim \Lambda)^2$, such that Wilson limits of order $10^{-4}$ may lead to energy limits of order 10 TeV, but this discussion is beyond the scope of the present article.

# 7. Summary

This article has reviewed the present status of neutron $\beta$ decay experiments with emphasis on new data for the neutron lifetime, the neutron $\beta$ asymmetry, the $\beta$-$\nu$ correlation, and the Fierz term. Their impact both within and beyond the standard model of particle physics is considerable, in particular on CKM unitarity, Equation 11, on putative dark neutron decays, Section 5.3.2, and on limits on tensor and other exotic couplings, Equation 13. Altogether, deviations from the V–A structure of the SM are excluded well below the $10^{-3}$ level. Fifteen years ago, these limits were still on the 10% level (8). New developments in EFT and lattice QCD theory have made it possible to compare constraints on New Physics via the appropriate Wilson coefficients from low- and high-energy experiments, see Table 5. It turns out that neutron and other $\beta$ decay experiments compare well with and are in part complementary to limits derived from LHC experiments.

## SUMMARY POINTS

1. Recent experiments on neutron $\beta$ decay gave strongly improved results.
2. These new results led to better values of basic Standard Model (SM) quantities, like the leading entry $V_{ud}$ of the CKM quark-mixing matrix, the axial coupling $g_A$, and the cross sections for neutrino-baryon reactions.
3. A main topic of contemporary particle physics is searches for New Physics beyond the SM. The new neutron and nuclear data permit exclusion of deviations from the V–A structure of the SM well below the $10^{-3}$ level, two orders of magnitude better than 15 years ago.

4. Deviations from the SM are parametrized by appropriate Wilson coefficients. Progress in effective field theory permits to compare limits on Wilson coefficients from neutron decay with corresponding limits from high-energy proton-proton collisions.
5. Limits on Wilson coefficients from low-energy experiments are generally more precise and require fewer assumptions than the corresponding high-energy limits.
6. High-energy experiments, by contrast, are more sensitive to non-SM right-handed neutrinos, and this higher sensitivity makes them complementary to the low-energy experiments.

## Acknowledgments

The authors acknowledge support by the Priority Programme SPP 1491 of the German Research Foundation (DFG). B.M. acknowledges support from the Excellence Cluster ORIGINS, which is funded by the DFG under Germany´s Excellence Strategy – EXC-2094 – 390783311.

## Terms and Definitions

**Acronyms of particle physics:**

| | |
|---|---|
| CKM | Cabibbo-Kobayashi-Maskawa quark-mixing |
| CL | Confidence Limit |
| CVC | Conserved Vector Current |
| EDM | Electric Dipole Moment |
| EFT | Effective Field Theory |
| LHC | Large Hadron Collider, CERN, Geneva, Switzerland |
| PDG | Particle Data Group |
| PCAC | Partially Conserved Axialvector Current |
| QED | Quantum ElectroDynamics |
| QCD | Quantum ChromoDynamics |
| QFT | Quantum Field Theory |
| SM | Standard Model |
| SMEFT | EFT at energies above the SM scale |

**Neutron sources used for neutron $\beta$ decay studies:**

**Reactor sources**

| | |
|---|---|
| FRM II | Forschungs-Neutronenquelle Maier-Leibnitz, Garching, Germany |
| ILL | Institut Max von Laue-Paul Langevin, Grenoble, France |
| NIST | National Institute of Standards and Technology, Gaithersburg, USA |

PNPI               Petersburg Nuclear Physics Institute, Russia

TRIGA-Mainz Mainz University, Germany

**Spallation sources**

ESS                European Spallation Source, Lund, Sweden, under construction

J-PARC           Japan Proton Accelerator Research Complex, Tokai, Japan

LANL             Los Alamos National Laboratory, USA

PSI                 Paul-Scherrer-Institut, Villigen, Switzerland

SNS                Spallation Neutron Source, Oak Ridge National Laboratory, USA

**Blinded Measurements:** To eliminate bias, separate teams of a collaboration analyze different parts of an experiment with some parameters hidden before being officially 'unblinded'.

# References

1. Abele H. *Prog. Part. Nucl. Phys.* 60:1 (2008)

2. Nico J. *J. Phys. G* 36:104001 (2009)

3. Dubbers D, Schmidt MG. *Rev. Mod. Phys.* 83:1111 (2011)

4. Wietfeldt FE, Greene GL. *Rev. Mod. Phys.* 83:1173 (2011)

5. Bhattacharya T, et al. *Phys. Rev. D* 85:054512 (2012)

6. Holstein BR. *J. Phys. G* 41:110301 (2014)

7. Jenke T, et al., editors. *EPJ Web Conf.* 219:00001 (2019)

8. Severijns N, Beck M, Naviliat-Cuncic O. *Rev. Mod. Phys.* 78:991 (2006)

9. Vos KK, Wilschut H, Timmermans RGE. *Rev. Mod. Phys.* 87:1483 (2015)

10. González-Alonso M, Naviliat-Cuncic O, Severijns N. *Prog. Part. Nucl. Phys.* 104:165 (2019)

11. Abel C, et al. *Phys. Rev. Lett.* 124:081803 (2020)

12. Baldo-Ceolin M, et al. *Z. Phys. C.* 63:405 (1994)

13. Berezhiani Z, et al. *Physics* 1:271 (2019)

14. Gardner S, Haxton WC, Holstein BR. *Annu. Rev. Nucl. Part. Sci.* 67:69 (2017)

15. Cronenberg G, et al. *Nature Physics* 14:1022 (2018)

16. Rauch H, Werner SA. *Neutron Interferometry*. 2015. Oxford: Oxford University Pres

17. Bitter T, Dubbers D. *Phys. Rev. Lett.* 59:251 (1987)

18. Muskat E, Dubbers D, Schärpf O. *Phys. Rev. Lett.* 58:2047 (1987)

19. Lushchikov VI, et al. *JETP Lett.* 9:23 (1969)

20. Steyerl A. *Phys. Lett. B* 29:33 (1969)

21. Steyerl A, et al. *Phys. Lett. A* 116:347 (1986)

22. Golub R, Pendlebury JM. *Phys. Lett. A* 53:133 (1975)

23. Bison G, et al. *Phys. Rev. C* 95:045503 (2017)

24. Maier-Leibnitz H, Springer T. *Annu. Rev. Nucl. Sci.* 16:207 (1966)
25. Häse H, et al. *Nucl. Instr. Meth. A* 485:453 (2002)
26. Abele H, et al. *Nucl. Instr. Meth. A* 562:407 (2006)
27. Mezei F. *Comm. Phys.* 1:81 (1976)
28. Serebrov AP, et al. *Phys. Lett. A* 309:218 (2003)
29. Ignatovich VK. *The Physics of Ultracold Neutrons*. 1990. Oxford: Oxford University Press
30. Golub R, Richardson D, Lamoreaux K. *Ultracold Neutrons.* 1991. Bristol: Hilger
31. Steyerl A. *Ultracold Neutrons.* 2020. Singapore: World Scientific
32. Schaerpf O. *Physica B* 156-157:639 (1989)
33. Petukhov AK, et al. *Rev. Sci. Instrum.* 90:085112 (2019)
34. Klauser C, et al. *Nucl. Instr. Meth. A* 840:181 (2016)
35. Mauri G, et al. *Eur. Phys. J. Tech. and Instr.* 6:3 (2019)
36. Köhli M, et al. *Nucl. Instr. Meth. A* 828:242 (2016)
37. Webber DM. *Phys. Rev. Lett.* 106:041803 (2011), 106:079901(E) (2011)
38. Cabibbo N, Swallow EC, Winston R. *Phys. Rev. Lett.* 10:531 (1963)
39. Kobayashi M, Maskawa T. *Progr. Theor. Phys.* 49:652 (1973)
40. Gell-Mann M. *Phys. Rev.* 111:362 (1958)
41. Goldberger ML, Treiman SB. *Phys. Rev.* 111:354 (1958)
42. Bali GS, et al. 05:126 (2020)
43. Ivanov AN, et al. *Phys. Rev. D* 88:073002 (2013)
44. Hayen L, et al. *Rev. Mod. Phys.* 90:015008 (2018)
45. Jackson JD, Treiman SB, Wyld Jr HW. *Phys. Rev.* 106:517 (1957)
46. Jackson JD, Treiman SB, Wyld Jr HW. *Nucl. Phys.* 4:206 (1957)
47. Ebel ME, Feldman G. *Nucl. Phys.* 4:2013 (1957)
48. Fierz M. *Z. Phys.* 104:553 (1937)
49. Glück F, Ioó I, Last J. *Nucl. Phys. A* 593:125 (1995)
50. Roick C. Doctoral thesis 2018, https://mediatum.ub.tum.de/doc/1452579/1452579.pdf
51. Weinberg S. *PoS (CD09)* 001 (2009)
52. Holstein BR. *Nucl. Phys. A* 689:135 (2001)
53. Lee TD, Yang CN. *Phys. Rev.* 105:1671 (1957)
54. Herczeg P. *Prog. Part. Nucl. Phys.* 46:413 (2001)
55. Cirigliano V, Gardner S, Holstein BR. *Prog. Part. Nucl. Phys.* 71:93 (2013)
56. Ando S, et al. *Phys. Lett. B* 595:250 (2004)
57. Marciano WJ, Sirlin A. *Phys. Rev. Lett.* 96:032002 (2006)
58. Willenbrock S, Zhang C. *Annu. Rev. Nucl. Part. Sci.* 64:83 (2014)
59. Buchmüller W, Wyler D. *Nucl. Phys. B* 268:621 (1986)
60. Grzadkowski B, et al. *J. High Energ. Phys.* 010:085 (2010)

61. Falkowski A, González-Alonso M, Mimouni K. *JHEP* 08:123 (2017)

62. Chang CC, et al. *Nature* 558:91 (2018)

63. Gupta R, et al. *Phys. Rev. D* 98:034503 (2018)

64. Liang J, et al. *Phys. Rev. D* 98:074505 (2018)

65. Harris T, et al. *Phys. Rev. D* 100:034513 (2019)

66. Aoki S, et al. *Eur. Phys. J. C* 80:113 (2020)

68. Cirigliano, et al. *J. High Energ. Phys.* [nucl-ex] (2019)

69. Zyla PA, et al. (Particle Data Group). *Prog. Theor. Exp. Phys.* 2020:083C01 (2020)

70. Robson JM, *Phys. Rev.* 83:349 (1951)

71. Christensen CJ, et al. *Phys. Rev. D* 5:1628 (1972)

72. Abov YG, et al. *JETP Letters* 44:6 (1986), http://www.jetpletters.ac.ru/ps/1393/article_21123.shtml

73. Paul W, et al. *Z. Phys. C* 45:25 (1989)

74. Mampe W, et al. *Nucl. Instr. Meth. A* 284:111 (1989)

75. Wietfeldt HE. *Atoms* 6:70 (2018)

76. Arzumanov S, et al. *Phys. Lett. B* 745:79 (2015)

77. Serebrov AP, et al. *Phys. Rev. C* 97:055503 (2018)

78. Fomin A, Serebrov A. *EPJ Web Conf.* 219:03001 (2019)

79. Serebrov, et al. *Phys.Lett.B* 605:72 (2005)

80. Dubbers D. *arXiv*:1807.07026 [hep-ph] (2018)

81. Morris CL, et al. *Rev. Sci. Instr.* 88:053508 (2017)

82. Pattie Jr RW, et al. *Science* 360:627 (2018)

83. Steyerl A, Gutsmiedl E. *Phys. Rev. C* 102:045203 (2020)

84. Ezhov VF, et al. *JETP Lett.* 107:671 (2018)

85. Yue AT, et al. *Phys. Rev. Lett.* 111:222501 (2013)

86. Pichlmaier A, et al. *Phys. Lett. B* 693:221 (2010)

87. Steyerl A, et al. *Phys. Rev. C* 85:0655023 (2012)

88. Materne S, et al. *Nucl. Instr. Meth. A* 611:176 (2009)

89. Leung KKH, et al. *Phys. Rev. C* 94:045502 (2016)

90. Karch JP. Doctoral thesis 2017, in German. https://d-nb.info/1143864549/34

91. Hirota K, et al. *Prog. Theor. Exp. Phys.* 2020:123C02 (2020)

92. Hoogerheide SF, et al. *EPJ Web Conf.* 219:03002 (2019)

93. Dubbers D. *Nucl. Instr. Meth. A* 1009:165456 (2021)

94. Roick C, et al. *Phys. Rev. C* 97:035502 (2018)

95. Ashtari Esfahani A, et al. *New J. Phys.* 22:033004 (2020)

96. Dubbers D, Schmidt U. *Nucl. Instr. Meth. A* 837:50 (2016)

97. Märkisch B, et al. *Phys. Rev. Lett.* 122:242501 (2019)

98. Saul H, et al. *Phys. Rev. Lett.* 125:112501 (2020)

99. Brown MA-P, et al. *Phys. Rev. C* 97:035505 (2018)

100. Beck M, et al. *Phys. Rev. C* 101:055506 (2020)

101. Otten EW, Weinheimer C. *Rep. Prog. Phys.* 71:086201 (2008)

102. Stratowa C, Dobrozemsky R, Weinzierl P. *Phys. Rev. D* 18:3970 (1978)

103. Byrne J, et al. *J. Phys. G* 28:1325 (2002)

104. Hassan MT, et al. *Phys. Rev. C* 103:045502 (2021)

105. Pattie RW, Hickerson KP, Young AR. *Phys. Rev. C* 88:048501 (2013)

106. Hickerson KP, et al. *Phys. Rev. C* 96:042501(R) (2017), 96:059901(E) (2017)

107. Sun X, et al. *Phys. Rev. C* 101:035503 (2020)

108. Serebrov AP, et al. *JETP* 86:1074, translated from ZETF 113:1963 (1998)

109. Kreuz M, et al. *Phys. Lett. B* 619:263 (2005)

110. Schumann M, et al. *Phys. Rev. Lett.* 99:191803 (2007)

111. Schumann M, et al. *Phys. Rev. Lett.* 100:151801 (2008)

112. Soldner T, et al. *Phys. Lett. B* 581:49 (2004)

113. Chupp TE, et al. *Phys. Rev. C* 86:035505 (2012)

114. Kozela A, et al. *Phys. Rev. C* 85:045501 (2012)

115. Bales MJ, et al. *Phys. Rev. Lett.* 116:242501 (2016)

116. Bernard V, et al. *Phys. Lett.* B 593:105 (2004)

117. Schott W, et al. *EPJ Web Conf.* 219:04006 (2019)

118. Počanić D, et al. *Nucl. Instr. Meth. A* 611:211 (2009)

119. Fry J, et al. *EPJ Web Conf.* 219:04002 (2019)

120. Dubbers D, et al. *Nucl. Instr. Meth. A* 596:238 (2008)

121. Wang X, et al. *EPJ Web Conf.* 219:04007 (2019)

122. Moser D, et al. *J. Phys. Conf. Ser.* 1643:012005 (2020)

123. Soldner T, et al. *EPJ Web Conf.* 219:10003 (2019)

124. Bodek K, et al. *EPJ Web Conf.* 219:04001 (2019)

125. Fields BD, et al. *JCAP03*(2020)010 (2020)

126. Towner IS, Hardy JC. *Rep. Prog. Phys.* 73:046301 (2010)

127. Seng CY, et al. *Phys. Rev. Lett.* 121:241804 (2018)

128. Feng X, et al. *Phys. Rev. Lett.* 124:192002 (2020)

129. Hardy JC, Towner IS. *Phys. Rev. C* 102:045501 (2020)

130. Gorchtein M. *Phys. Rev. Lett.* 123:042503 (2019)

131. Seng CY, Gorchtein M, Ramsey-Musolf MJ. *Phys. Rev. D* 100:013001 (2019)

132. Czarnecki A, Marciano WJ, Sirlin A. *Phys. Rev. D (R)* 101:091301 (2020)

133. Falkowski A, González-Alonso M, Naviliat-Cuncic O. .*J. High Energy Phys.* 2104:126 (2021)

134. De Florian, D, et al. https://e-publishing.cern.ch/index.php/CYRM/issue/view/32/3 (2017)

135. Seng CY, et al. *Phys. Rev. D* 101:111301(R) (2020)
136. Fornal B, Grinstein B. *Phys. Rev. Lett.* 120:191801 (2018)
137. Dubbers D, et al. *Phys. Lett. B* 791:6 (2019)
138. Czarnecki A, Marciano WJ, Sirlin A. *Phys. Rev. Lett.* 120:202002 (2018)
139. Hayen L, Young AR. *arXiv*:2009.11364 [nucl-ex] (2020)
140. Combs D, et al. *arXiv*:2009.13700 [nucl-ex] (2020)
141. Ando S, McGovern JA, Sato T. *Phys. Lett. B* 677:109 (2009)
142. Beg MAB, et al. *Phys. Rev. Lett.* 38:1252 (1977)
143. Severijns N. *J. Phys. G: Nucl. Part. Phys.* 41:114006 (2014)
144. Collady D, Kostelecký VA. *Phys. Rev. D* 58:116002 (1998)
145. Nițescu O, Ghinescu S, Stoica S. *J. Phys. G: Nucl. Part. Phys.* 47**:**055112 (2020)
146. Bodek K, et al. PoS(X LASNPA)029 (2014)
147. Crivellin A, Hoferichter M. Phys. Rev. Lett. 125:111801 (2020)
148. Roberts BM, Dzuba VA, Flambaum VV. *Annu. Rev. Nucl. Part. Sci.* 65:63 (2015)
149. Khachatryan V, et al. *Phys. Rev. D* 91:092005 (2015)
150. Aad G, et al. *J. High Energ. Phys.* 08:009 (2016)
151. Aad G, et al. *J. High Energ. Phys.* 2020:5 (2020)